\documentclass{elsart3}

\usepackage{epsfig}

\usepackage{dcolumn}
\usepackage{bm}
\usepackage{graphicx}
\usepackage{amssymb}

\begin{document}

\begin{frontmatter}

\title{The smallest free-electron sphere sustaining multipolar surface plasmon
oscillation}
\author{K. Kolwas, A. Derkachova and S. Demianiuk}
\address {Institute of Physics of the Polish Academy of Sciences
Al.Lotnik\'{o}w 32/46, 02-668 Warsaw, Poland}
\date{}
\maketitle

\begin{abstract}
We study the oscillation frequencies and radiative decay rates of surface
plasmon modes of a simple-metal sphere as a function of sphere radius
without any assumptions concerning the sphere size. We re-examine within the
framework of classical electrodynamics the usual expectations for multipolar
plasmon frequency in the so called ''low radius limit'' of the classical
picture.
\end{abstract}


\end{frontmatter}

\section{Introduction}

The dielectric properties of metals, as well as those of semiconductors with
high electron concentration, are due to collective effects arising from the
Coulomb interaction between charges. In simple bulk metals the conduction
electrons can be considered as a free-electron plasma. Frequency dependence
of some of optical properties can be well described at a quantitative level
by the Drude-Lorentz dielectric function ~\cite{kittel}. Optical properties
of the electron gas in bulk metals, in proximity semi-infinitive surfaces,
in thin films, and in metallic particles can be characterized by the
eigenfrequencies of the system depending on free electron density and the
geometry of the system. If we talk of ''plasmons'' or ''plasma waves'', we
mean eigenmodes of the self-consistent Maxwell equations for the system in
the absence of an external electromagnetic field (or in a direction
orthogonal to the field) (e.g. \cite{reather}, \cite{forstmann}). ''Surface
plasmons'', are used as a name for electromagnetic eigenmodes which are
maximal near the surface. The time dependence of eigenmodes of a
free-electron system is characterized by corresponding eigenfrequencies with
the real part defining the frequency of oscillation, and the imaginary part
defining the radiative damping.

Usually the eigenmode problem of a metallic sphere is studied in the limit
of very small size parameter (retardation effect omitted), e.g. \cite{b and
h}, \cite{forstmann}, \cite{rupin} and the radiative damping of plasmon
oscillation is not included. The dipole mode eigenfrequency is then expected
to be equal to $\omega _{p}/\sqrt{3}$ being responsible for the ''giant
dipole resonance'' resulting from the Mie scattering theory (\cite{mie}, and
also e.g. \cite{b and w}, \cite{b and h}, \cite{stratton}, \cite{kreibig}). $%
\omega _{p}$ is the modified plasma frequency, which can include or not the
cluster core polarizability and (or) the spill-out effect of electron
density at particle border, depending on the model approximations in effect.

In \cite{kolwas1}, \cite{kolwas2} we have reconsidered the eigenvalue
problem of a free-electron metal sphere as a function of sphere radius
without any assumption concerning the particle size and including higher
eigenmodes than the dipole ones. We have studied the dipole ($l=1$) and the
higher polarity plasmon eigenfrequencies $\omega _{l}(R)$ as well as the
plasmon radiative decay rates $\omega _{l}^{\prime \prime }(R)$ as a
function of the particle radius $R$ with no assumption concerning the lower
limit of the particle size in numerical modelling (retardation effects
included) for $l=1,2,...6$. In \cite{kolwas2} we have also studied the
plasmon manifestation in scattering and absorbing properties of the sphere
of arbitrarily large size (retardation included) within full scattering Mie
theory.

In the present paper, we use the same ''exact'' solutions of the eigenmode
problem for $l=1,2,...6$ and $7,8...10$ in addition, and re-examine the
usual expectation for multipolar plasmon frequencies in the so called ''low
radius limit''. If the particle is formed from ideal metal (free electrons
do not suffer from collisions $\gamma =0$) and is embedded in vacuum ($%
\varepsilon _{out}=1$) the multipolar plasmon frequencies according to the
''low radius limit'' approximation are expected to be $\omega _{0,l}=\omega
_{p}\sqrt{l/\left( 2l+1\right) }$ (e.g. \cite{ritchie}, \cite{boardman},
\cite{forstmann}, \cite{rupin}). The well known dipole mode frequency $%
\omega _{p}/\sqrt{3}$ is obtained for $l=1$, while for increasing $l$ the
eigenmode frequencies approach the frequency of plane surface plasmon at $%
\omega _{p}/\sqrt{2}$, in spite of the fact they result from the ''low
radius approximation'' (i.e. from the limit of $R\rightarrow 0$, while plane
surface limit is $R\rightarrow \infty $). In this paper we study the reasons
of underlying causes for this paradox.

\section{Formulation of the eigenvalue problem for a sphere of arbitrary size%
}

The starting point is provided by the self-consistent Maxwell equations:

\begin{equation}
\begin{array}{cc}
\nabla \times \mathbf{B}=\frac{1}{c^{2}}\frac{\partial \mathbf{E}}{\partial t%
}+\mu _{0}\mathbf{j} & \ \ \ \ \ \nabla \cdot \mathbf{E}=\rho /\varepsilon
_{0} \\
\nabla \times \mathbf{E}=-\frac{\partial \mathbf{B}}{\partial t} & \ \ \
\nabla \cdot \mathbf{B}=0
\end{array}
\end{equation}

with no external sources: $\rho _{ext}=0$,$\ \mathbf{j}_{ext}=0$ so $\mathbf{%
j}$ and $\rho $ are induced current and charge densities respectively.

The frequency dependent dielectric function $\varepsilon (\omega
)=\varepsilon _{in}(\omega )$ and conductivity $\sigma (\omega )=\sigma
_{in}(\omega )$ of the sphere is assumed to have the constant bulk value up
to the sphere border. The dynamic, linear response of the sphere material is
described within standard optics, so the local proportionality between the
electric displacement $\mathbf{D}$ and electric field intensity $\mathbf{%
\mathbf{E}}$ at the same point in space are valid: $\mathbf{D(\mathbf{r}%
,\omega )=\mathbf{E}(\mathbf{r},\omega )}+\frac{i}{\omega }\mathbf{j}(%
\mathbf{r},\omega )=\varepsilon _{0}(1+\frac{i\sigma (\omega )}{\varepsilon
_{0}\omega })\mathbf{E}(\mathbf{r},\omega )\mathbf{=}\varepsilon
_{0}\varepsilon (\omega )\mathbf{E}(\mathbf{r},\omega )$. The sphere is
embedded in nonconducting and nonmagnetic medium $\sigma _{out}=0$ and $%
\varepsilon (\omega )=\varepsilon _{out}$\ will be assumed to be $%
\varepsilon _{out}=1$ in all numerical illustrations. The dielectric
function of the sphere will be assumed to be the Drude dielectric function $%
\varepsilon _{in}(\omega )=1-\frac{\omega _{p}^{2}}{\omega ^{2}-i\gamma
\omega }$. We look for solutions fulfilling Maxwell's equations in the form
of transversal waves ($\nabla \cdot \mathbf{E}$ $=0$) in two homogeneous
regions inside and outside the sphere so the wave equation: $\nabla ^{2}%
\mathbf{E}(\mathbf{r})+\mathbf{\nabla }(\mathbf{\nabla }\cdot \mathbf{E}(%
\mathbf{r}))-\frac{1}{c^{2}}\frac{\partial ^{2}\mathbf{D}}{\partial t^{2}}=0$
for harmonic fields $\mathbf{\mathbf{E}(\mathbf{r},\omega )=\mathbf{E}}%
e^{-i(kr-\omega t)}$ reduces to the Helmholtz equation:

\begin{equation}
\nabla ^{2}\mathbf{E}(\mathbf{r})+q^{2}\mathbf{E}(\mathbf{r})=0  \label{Weq}
\end{equation}

where: $q=q_{in}$ inside the sphere, $q=q_{out}$ in the sphere surroundings,
$q_{in}=q_{0}\sqrt{\varepsilon _{in}}$ , $q_{out}=q_{0}\sqrt{\varepsilon
_{out}}$ and $q_{0}=\frac{\omega }{c}$. The well known scalar solution of
the corresponding scalar equation (e.g.\cite{b and h}, \cite{stratton}) in
spherical coordinates ($r,\theta ,\phi $) reads:

\begin{equation}
\psi _{lm}(r,\theta ,\phi )=Z_{l}(qr)Y_{lm}(\theta ,\phi ),
\label{ScallarEq}
\end{equation}

where $l=1,2,...,$\ $m=0,\pm 1,...,\pm l$, $Y_{lm}(\theta ,\phi )$ are
spherical harmonics, and $Z_{l}(qr)$ are spherical Bessel functions $%
j_{l}(q_{in}r)$ inside the sphere and the spherical Hankel functions $%
h_{l}(q_{out}r)$ outside the sphere.

Because various notations have been employed in different papers and
textbooks and none appears to have general acceptance, let's recall that the
spherical Bessel functions: $j_{l}\left( z\right) =\sqrt{\frac{\pi }{2z}}%
J_{l+\frac{1}{2}}(z),$ and $h_{l}\left( z\right) =j_{l}\left( z\right)
-i\cdot n_{l}(z)=\sqrt{\frac{\pi }{2z}H_{l+\frac{1}{2}}^{(1)}(z)}$ where $%
n_{l}(z)=\sqrt{\frac{\pi }{2z}}N_{l+\frac{1}{2}}(z)$. The functions $J_{l+%
\frac{1}{2}}(z)$, $H_{l+\frac{1}{2}}^{\left( 1\right) }(z)$ and $N_{l+\frac{1%
}{2}}(z)$ are Bessel, Hankel and Neuman cylindrical functions of half order
of the standard type according to the convention used e.g. in \cite{b and w}.

From scalar solution $\psi _{lm}$ one can construct two independent
solutions of the vectorial wave equation (\ref{Weq}), one with vanishing
radial component of the magnetic field:

\begin{eqnarray}
\mathbf{E}(\mathbf{r}) &=&B_{lm}(1/q)\mathbf{\nabla \times \nabla \times (r}%
\psi _{lm}\mathbf{),} \\
\mathbf{H}(\mathbf{r}) &=&B_{lm}(q/iq_{0})\mathbf{\nabla \times (r}\psi _{lm}%
\mathbf{).}
\end{eqnarray}

and the other with vanishing radial component of the electric field:

\begin{eqnarray}
\mathbf{E}(\mathbf{r}) &=&A_{lm}\mathbf{\nabla \times (r}\psi _{lm}\mathbf{),%
}  \label{TEa} \\
\mathbf{H}(\mathbf{r}) &=&A_{lm}(1/q_{0})\mathbf{\nabla \times \nabla \times
(r}\psi _{lm}\mathbf{),}  \label{TEb}
\end{eqnarray}

$A_{lm}$ and $B_{lm}$ are constants that take different values $A_{lm}^{in}$
and $B_{lm}^{in}$ inside and $A_{lm}^{out}$ and $B_{lm}^{out}$ outside the
sphere. The explicit expressions for the solution with the nonzero radial
component of the electric field $E_{r}\neq 0$ (and the magnetic field
tangent to the sphere surface $H_{r}=0)$, which is named transverse magnetic
(TM) mode in analogy to the flat surface interface case (\emph{p}
polarization, or ''electric wave'' in terminology of \cite{b and w}) read:

\begin{eqnarray}
E_{r}(r,\theta ,\phi )
&=&B_{lm}l(l+1)(qr)^{-1}Z_{l}(qr)Y_{lm}(\theta ,\phi
),  \nonumber \\
E_{\theta }(r,\theta ,\phi )
&=&B_{lm}(qr)^{-1}[qrZ_{l}(qr)]^{\prime
}\partial Y_{lm}/\partial \theta ,  \nonumber \\
E_{\varphi }(r,\theta ,\phi ) &=&B_{lm}im(qr\sin \theta
)^{-1}[qrZ_{l}(qr)]^{\prime }Y_{lm}(\theta ,\phi ),  \nonumber \\
H_{r}(r,\theta ,\phi ) &=&0,  \label{TM1} \\
H_{\theta }(r,\theta ,\phi ) &=&B_{lm}[\varepsilon (\omega
)]^{1/2}(m/\sin
\theta )Z_{l}(qr)Y_{lm}(\theta ,\phi ),  \nonumber \\
H_{\varphi }(r,\theta ,\phi ) &=&iB_{lm}[\varepsilon (\omega
)]^{1/2}Z_{l}(qr)\partial Y_{lm}/\partial \theta ,  \nonumber
\end{eqnarray}

The expression for the orthogonal solution with $E_{r}=0$ results from eqs.(%
\ref{TEa},\ref{TEb}) (and is named transverse electric (TE) mode in analogy
to the flat surface interface case (\emph{s} polarization)). The prime
indicates differentiation in respect to the argument, which is $q_{in}r$ or $%
q_{out}r$ correspondingly. We focus our attention on TM\ mode only.

The continuity relations at the sphere boundary for the tangential
components of the electric field (the continuity of $E_{\theta }$ and $%
E_{\varphi }$)\ lead to the same condition:

\begin{equation}
B_{lm}^{in}(z_{B})^{-1}[z_{B}j_{l}(z_{B})]^{\prime
}=B_{lm}^{out}(z_{H})^{-1}[z_{H}h_{l}(z_{H})]^{\prime }
\end{equation}

while the tangential components of the magnetic field (the continuity of $%
H_{\theta }$ and $H_{\varphi }$) lead to the condition:

\begin{equation}
B_{lm}^{in}\sqrt{\varepsilon _{in}}j_{l}(z_{B})=B_{lm}^{out}\sqrt{%
\varepsilon _{out}}h_{l}(z_{H})
\end{equation}

where:

\begin{equation}
z_{B}=q_{in}R=\frac{\omega }{c}R\sqrt{\varepsilon _{in}}  \label{zB}
\end{equation}

is the argument of the Bessel function $j_{l}$, and

\begin{equation}
z_{H}=q_{out}R=\frac{\omega }{c}R\sqrt{\varepsilon _{out}}=z_{B}\frac{\sqrt{%
\varepsilon _{out}}}{\sqrt{\varepsilon _{in}}}  \label{zH}
\end{equation}

is the argument of the Hankel function for $r=R$. The continuity relations
for TM mode lead to non-trivial solutions (e.g. non-zero field amplitudes $%
B_{lm}$ inside and outside the sphere) only when:

\begin{equation}
\frac{\lbrack z_{B}j_{l}(z_{B})]^{\prime }}{\varepsilon _{in}j_{l}(z_{B})}=%
\frac{[z_{H}h_{l}(z_{H})]^{\prime }}{\varepsilon _{out}h_{l}(z_{H})}
\label{TMDR}
\end{equation}

We are interested in the properties of the sphere in the frequency regime of
anomalous dispersion $\varepsilon _{in}(\omega )<0$. In that region only the
TM eigenmodes exist, while the equation dispersion relation for TE mode has
no solution for $\varepsilon _{in}(\omega )<0$ (\cite{rupin} or \cite{halevi}%
). $Z_{l}(qr)=j_{l}(q_{in}r)$ is then a function of a complex
argument and the solutions given by eqs. (\ref{TM1}) are called
''surface modes''. The fields are maximal at the sphere surface,
with exception of the $l=1$ mode which is uniform throughout the
sphere (\cite{b and h} or \cite {halevi}).

On writing down the dispersion relation for the TM mode (\ref{TMDR}) in
terms of the more compact Riccati-Bessel function $\psi _{l}\left( z\right)
=z\cdot j_{l}(z)$ and $\xi _{l}\left( z\right) =z\cdot h_{l}^{(1)}(z)$, the
dispersion relation for the TM mode reads:

\begin{equation}
\sqrt{\varepsilon _{out}}\xi _{l}\left( z_{H}\right) \psi _{l}^{\prime
}\left( z_{B}\right) -\sqrt{\varepsilon _{in}}\psi _{l}\left( z_{B}\right)
\xi _{l}^{\prime }\left( z_{H}\right) =0  \label{RD}
\end{equation}

The boundary conditions are then satisfied only by a discrete set of
characteristic complex values $z_{l}$ which are the roots of the complex
function $D_{l}(z)\equiv \sqrt{\varepsilon _{out}}\xi _{l}\left(
z_{H}(\omega )\right) \psi _{l}^{\prime }\left( z_{B}(\omega )\right) -\sqrt{%
\varepsilon _{in}(\omega )}\psi _{l}\left( z_{B}(\omega )\right) \xi
_{l}^{\prime }\left( z_{H}(\omega )\right) $ of complex argument $z=z(\omega
,R)$. Discretization of complex roots $z_{l}$ means the discretization of
corresponding values $\omega =\Omega _{l}$, $l=1,2,3...$ which are allowed
to be complex: $\Omega _{l}=\omega _{l}+i\omega _{l}^{\prime \prime } $.
They define discrete eigenmode frequencies $\omega _{l}$ and damping rates $%
\omega _{l}^{\prime \prime }$ for the TM mode being the sum of
corresponding components of (\ref{TM1}) multiplied by $e^{i\Omega
_{l}t}=e^{i\omega _{l}t}e^{\omega _{l}^{\prime \prime }t}$. The
analytic form of $z_{l}=z_{l}(\Omega _{l}(R),R)$ is not known, nor
the analytic form of the relation $\Omega _{l}(R)$. Let's notice,
that neither $z_{H}(\omega )$ nor $z_{B}(\omega )$ separately are
appropriate to define the set of discrete characteristic values,
contrary to what is suggested in \cite {stratton}.

We solved the dispersion relation (\ref{RD}) with respect to $\Omega _{l}$
numerically by treating the radius $R$ as an external parameter.
Riccati-Bessel functions $\psi _{l}$, $\chi _{l}$ and $\xi _{l}$ (and their
derivatives with respect to the corresponding arguments $z_{H}$ and $z_{B}$)
were calculated exactly with use of the recurrence relation.

\begin{figure}[t]
\centerline{\includegraphics[height=60mm]{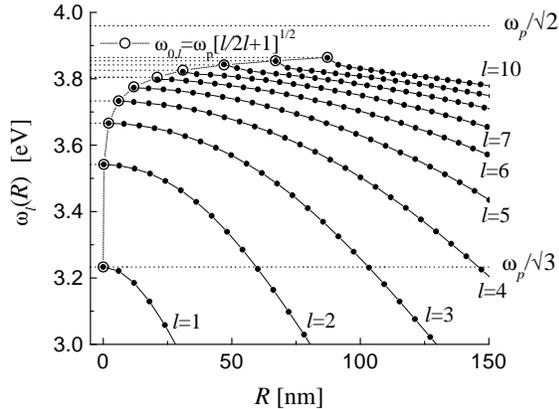}}
\caption{Plasmon oscillation frequencies $\protect\omega _{l}(R)$
as a function of sodium sphere radius $R$ for $l=1,2,...10$
(rigorous solution). The picture illustrates the coincidence of
the plasmon frequencies $\protect\omega _{l}(R_{\min ,l})$ with
the corresponding value $\protect\omega _{0,l}$ obtained within
vanishing size approximation (open circles). $\protect\gamma =0$.
}
\end{figure}

\begin{figure}[t]
\centerline{
\includegraphics[height=60mm]{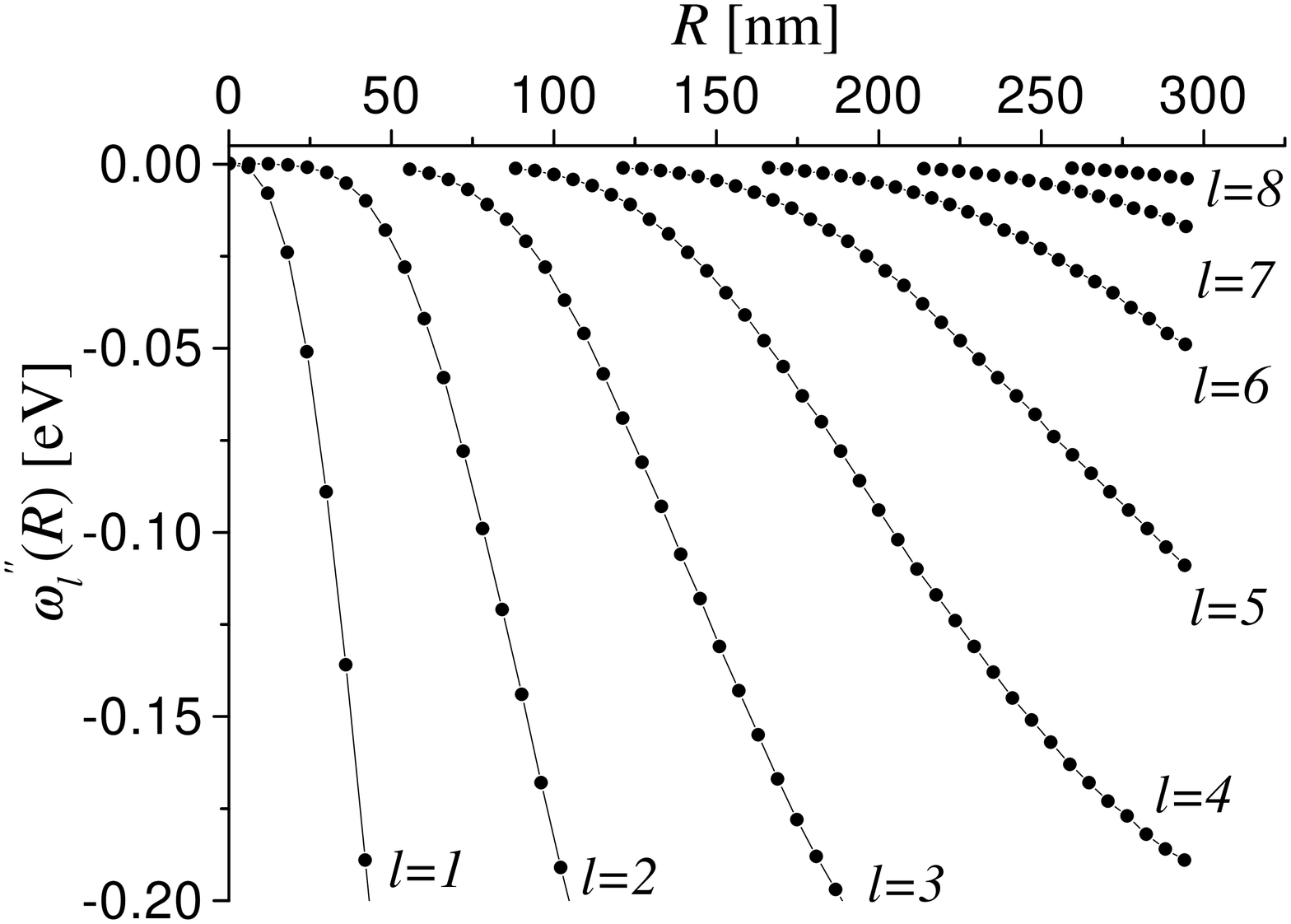}
           }
\caption{Rate of plasmon radiative damping $\protect\omega
_{l}^{\prime \prime }(R)$ as a function of sodium sphere radius
$R$ for $l=1,2,...10$ resulting from non-approximated radius
dependence for $\gamma =0$. }
\end{figure}

We have used the Mueller method of secants of finding numerical
solutions of the function $f(v)=0$ when one knows the starting
approximated values lying in the vicinity of the exact function
parameter $v,$which can be complex (the ''root'' function of the
Mathcad program). For given $l$ and given $R$, the complex
eigenvalue $\Omega _{l}$ was treated as the parameter to find,
successive values of $R$ were external parameters and where
changed with the step $\Delta R\approx 2$nm up to the final radius
value $R=300$nm. The values for $\omega _{l}(R)$ and $\omega
_{l}^{\prime \prime }(R)$ were searched for by starting from
approximate values of the root procedure chosen from the range
from $\omega _{p}\sqrt{3}$ up to $\omega _{p}\sqrt{2}$
correspondingly and for negative values of $\omega _{l}^{\prime
\prime }$. The numerical illustrations have been made for a sodium
sphere described by the Drude dielectric function with $\omega
_{p}=5.6$ eV.

\section{Results}

Very careful study of roots of the function $D_{l}(\Omega _{l})$
of
parameter $\Omega _{l}(R)$ for given $l$ for the decreasing limit of radii $%
R $ leads to the conclusion, that if the sphere is of the radius
smaller than the characteristic radius $R_{\min ,l}$, there exist
no $\Omega _{l}(R)$ real nor complex. So the complex
eigenfrequencies $\Omega _{l}(R)=\omega _{l}(R)+i\omega
_{l}^{\prime \prime }(R)\ $can be attributed to the sphere
starting from the characteristic radius $R=R_{\min ,l}\neq 0$ in
given $l$. There exist no purely real solution for $\Omega _{l}$:
surface plasmons are always damped, even if the dielectric
function $\varepsilon (\omega )$ is real ($\gamma =0$).

\begin{figure}[t]
\centerline{\includegraphics[height=55mm]{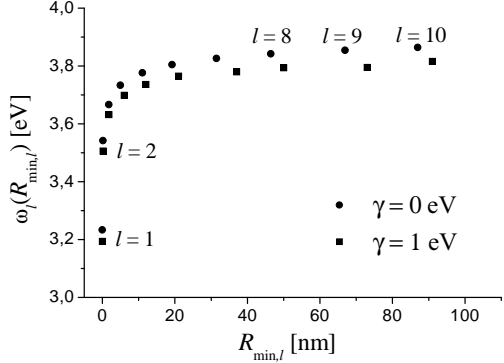}}
\caption{The relation of the plasmon frequency $\protect\omega _{l}$ and $%
R_{\min ,l}$ for successive values of $l=1,2,...10\ $ for electron
relaxation rates $\protect\gamma =0$ and $\protect\gamma =1$ eV}
\end{figure}

\begin{figure}[t]
\centerline{\includegraphics[height=55mm]{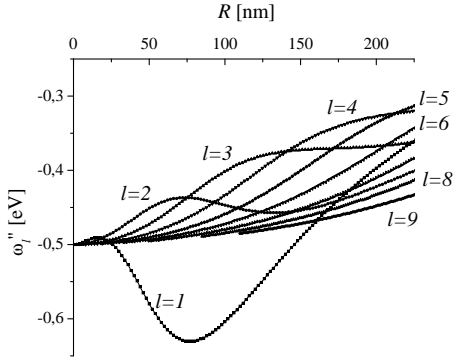}} \caption{The
dependence of damping rates: $\omega _{l}^{\prime \prime }(R)$ for
the electron relaxation rate $\protect\gamma =1$ eV for successive
values of $l=1,2,...10$.}
\end{figure}

Figure 1 and 2 (solid lines with closed spheres) illustrate the obtained $%
\omega _{l}(R)$ and $\omega _{l}^{\prime \prime }(R)$ dependencies for $%
\gamma =0$ and $l=1,2,3,...10$ starting from $\omega _{l}(R_{\min ,l})$ and $%
\omega _{l}^{\prime \prime }(R_{\min ,l})$ values. These figures complete
the picture for the $R\rightarrow R_{\min ,l}$ limit of the corresponding
dependence presented in \cite{kolwas1}, \cite{kolwas2} for $l=1,2,...6$,
figures 1 and 3, while in \cite{kolwas1} we did not study the limiting case
of $\omega _{l}(R\rightarrow R_{\min ,l})$ nor $\omega _{l}^{\prime \prime
}(R\rightarrow R_{\min ,l})$ in detail. More careful search for these
frequencies in the limit of smallest sphere still characterized by the
eigenvalues $\omega _{l}(R_{\min ,l})$ have shown, that they tend to the
values which can be approximated by $\omega _{0,l}$ values:

\begin{equation}
\omega _{l}(R_{\min ,l})\approx \omega _{0,l}=\omega _{p}\sqrt{\frac{l}{2l+1}%
},  \label{omega0}
\end{equation}

as illustrated by the hollow circles in figure 1. Our numerical experiment
shows that $R_{\min ,l}$ dependence on $l$ can be described as $R_{\min
,l}\approx C\left[ l\left( 2l+1\right) \right] ^{3/2}$ with the
proportionality constant $C$ depending on density of free electrons. $%
R_{\min ,l}$ can be e.g.: $R_{\min ,l=4}=6$nm, but it can be as large as $%
R_{\min ,l=10}=87.2$nm (the size parameter $2\pi R/\lambda \simeq 1$ for
optical wavelength $\lambda $).

The frequencies $\omega _{0,l}$ result from the dispersion relation (\ref{RD}%
) in the limit of small size parameter of the power series expansion of the
spherical Bessel and Hankel functions.

\begin{eqnarray}
j_{l}(z) &=&\frac{z^{l}}{(2l+1)!!}\left[ 1-\frac{0.5z^{2}}{1!(2l+3)}+ ....%
\right] \\
h_{l}(z) &=&-i\frac{(2l-1)!!}{z^{l+1}}\left[ 1-\frac{0.5z^{2}}{1!(1-2l)}+
....\right]
\end{eqnarray}

where $(2l\pm 1)!!\equiv 1\times 3\times 5\times ...\times (2l\pm 1)$. If
one employs the widely used rough approximation (e.g.\cite{forstmann}, \cite
{rupin}, \cite{halevi}):

\begin{eqnarray}
\psi _{l}(z_{B}) &\simeq &\widetilde{\psi }_{l}(z_{B})=\frac{z_{B}^{l+1}}{%
(2l+1)!!}, \\
\xi _{l}(z_{H}) &\simeq &\widetilde{\xi }(z_{H})=-i\frac{(2l-1)!!}{z_{H}^{l}}%
,
\end{eqnarray}

the dispersion relation (\ref{RD}) is fulfilled for any radius $R$ of the
sphere, and leads to the relation:

\begin{equation}
-\frac{l}{l+1}\frac{\varepsilon _{in}(\omega )}{\varepsilon _{out}}=1
\end{equation}

giving discrete plasmon frequencies:

\begin{equation}
\omega _{0,l}=\omega _{p}\sqrt{\frac{l}{2l+1}}
\end{equation}

which are real, in contrary to the exact solutions presented in figures 1
and 2 which are obligatory complex.

$\omega _{l}(R)$ dependence resulting from the exact solution do not
smoothly tend to the value $\omega _{l}(R\rightarrow 0)$ with decreasing $R$%
, as usually expected (e.g.\cite{forstmann}, \cite{rupin}, \cite{kolwas1}),
but it grows up to $\omega _{l}(R_{\min ,l})$ value, as illustrated in
figures 1. For $R<$ $R_{\min ,l}$ there are no eigenvalues $\Omega _{l}(R)$.
This behavior of the $\Omega _{l}(R)$ dependence is mainly due to fast
divergence of the $\xi _{l}(z_{H})$ function entering the dispersion
relation (\ref{RD}) for the arguments smaller than the range of variability
of $z_{H}=z_{H}(\Omega _{l}(R),R)$ parameters for successive $l$.

When one includes the relaxation rate of the electron gas into the
Drude model of the dielectric function, the plasmon frequency
$\omega _{l}$ for given radius $R$ of the sphere is relatively
slightly red shifted, while $\omega _{l}^{\prime \prime }$
experiences strong modification as illustrated in figure 3 and 4
respectively for $\gamma =1$ eV (\cite{kolwas4}).

\section{Conclusions}

By carefully studying the radius dependence of eigenmode problem of a sphere
one can formulate several conclusions allowing for better understanding of
surface plasmon features. In this paper we concentrate on studying the
differences of surface plasmon features in the classical picture resulting
from treating the radius dependence exactly, and the expectations from the
widely applied approximation of the so called ''low radius limit''. We use
the example of sodium sphere of plasma frequency $\omega _{p}=5.6$ eV ,
however the conclusions are qualitatively valid for other simple
free-electron metals. According to the non-approximated treatment the
surface plasmons are always radiatively damped, even in the absence of
collisional process: eigenfrequencies must be complex. The ''low radius
limit'' leads to the real eigenfrequencies $\omega _{0,l}$, which are radius
independent. From the exact calculations one can conclude, that the radius
dependence of multipolar plasmon frequencies is more subtle, than expected.
Our calculations show, that at larger polarity the $\omega _{l}(R)$
dependence does not smoothly tend to the value $\omega _{0,l}=\omega _{p}%
\sqrt{l/\left( 2l+1\right) }$\ of the vanishing size limit, as one could
expect (e.g.\cite{ritchie}, \cite{boardman}). If the sphere is of radius $R$
smaller than the characteristic radius $R_{\min ,l}\sim \left[ l\left(
2l+1\right) \right] ^{3/2}$, there is no\ related eigenvalue $\Omega _{l}(R)$
real nor complex. So the complex eigenfrequencies $\Omega _{l}(R)=\omega
_{l}(R)+i\omega _{l}^{\prime \prime }(R)\ $can be attributed to the sphere
starting from the radius $R=R_{\min ,l}\neq 0$. The radii $R_{\min ,l}$ for
higher polarities $l$ are not much smaller then the wavelength of the
optical range (the anomalous dispersion range of alkalies) so the ''low
limit approximation'' loses its validity. Our ''numerical experiment''
proves, that for the smallest particle radius $R_{\min ,l}$ still possessing
an eigenfrequency in given polarity $l$, the plasmon oscillation frequencies
can be well approximated by the corresponding value resulting from the ''low
radius limit'' approximation: $\omega _{l}(R_{\min ,l})\approx \omega _{p}%
\sqrt{l/\left( 2l+1\right) }$. Even though the problem of the optical
properties of metal sphere is at least as old as Mie theory \cite{mie}, it
seems, that the limitation for the smallest cluster still enabling the
plasmon oscillations has not been discussed previously.

\bigskip

This work was partially supported by the Polish State Committee for
Scientific Research (KBN), grant No. 2 P03B 102 22.\vspace*{0.04in}\newline

\end{document}